\documentclass[]{article}

\usepackage{amssymb}
\usepackage{color}
\usepackage{units}
\usepackage{graphicx}
\usepackage[round]{natbib}
\usepackage{hyperref}
\usepackage[T1]{fontenc}

\def\abs#1{\left|#1\right|}
\def\iu{\mathbf{i}}
\def\rmx{\mathrm{x}}
\def\rmy{\mathrm{y}}
\def\rmxy{\mathrm{xy}}

\def\vecstyle{\mathbf}
\def\matstyle{\mathbf}
\def\opstyle{\mathrm}

\title{Bias-Free Estimation of the Auto- and Cross-Covariance and the Corresponding Power Spectral Densities from Gappy Data}
\author{Nils~Damaschke\\[1.5mm]
Institute of General Electrical Engineering, University of Rostock\\Faculty of Computer Science and Electrical Engineering\\Albert-Einstein-Stra\ss{}e 2, 18059 Rostock, Germany\\[1.5mm]
nils.damaschke@uni-rostock.de\\[10mm]
Volker~K\"uhn\\[1.5mm]
Institute of Communications Engineering, University of Rostock\\Faculty of Computer Science and Electrical Engineering\\Richard-Wagner-Stra\ss{}e 31, 18119 Rostock-Warnem\"unde, Germany\\[1.5mm]
volker.kuehn@uni-rostock.de\\[10mm]
Holger~Nobach\\[1.5mm]
Max Planck Institute for Dynamics and Self-Organization\\Am Fa\ss{}berg 17, 37077 G\"ottingen, Germany\\[1.5mm]
holger.nobach@nambis.de}

\begin{document}

\maketitle

\begin{abstract}
Signal processing of uniformly spaced data from stationary stochastic processes with missing samples is investigated.
Besides randomly and independently occurring outliers also correlated data gaps are investigated.
Non-parametric estimators for the mean value, the signal variance, the autocovariance and cross-covariance functions and the corresponding power spectral densities are given, which are bias-free, independent of the spectral composition of the data gaps.
Bias-free estimation is obtained by averaging over valid samples only from the data set.
The procedures abstain from interpolation of missing samples.
An appropriate bias correction is used for cases where the estimated mean value is subtracted out from the data.
Spectral estimates are obtained from covariance functions using Wiener-Khinchin's theorem.
\end{abstract}

\maketitle

%
%

\section{Introduction}

In normal operation, measurement instruments usually deliver a continuous stream of equidistantly spaced samples of an observed, physical quantity.
Various reasons exist, which lead the normal operation to fail.
There might be boundary conditions, under which the measurement system cannot operate.
This includes cases, where the quantity under observation temporarily is not accessible.
For distributed measurement systems, the communication channels might be temporarily disturbed.
Also re-configuration or maintenance interruption might be necessary for the instrument.
In many cases, the measurement principle includes a signal pre-processing, which may fail under certain conditions and lead to outliers or gaps in the data stream.

Unfortunately, signal processing algorithms for statistical analysis usually require a continuous stream of equidistantly sampled data, typically arranged in blocks of a defined duration.
A widely used way to circumvent the problem caused by missing samples is to interpolate the signal and fill the gaps with predicted values, see \citet{vio_etal_00}.
The interpolation scheme can mime the statistical properties obtained from the valid parts of the signal and missing values can be predicted. 
Then the statistical quantities like the mean value, the variance, the covariance function or the power spectral density are derived from the reconstructed signal consisting of a mixture of originally valid samples and the interpolated ones.
Examples for this principle are Kalman interpolation in audio reconstruction by \citet{niedzwiecki_cisowski_93,preihs_etal_12}, the adaptive filter-bank approach by \citet{stoica_etal_00} or the Karhunen-Lo\`eve procedure resp.\ proper orthogonal decomposition for gappy data by \citet{everson_sirovich_95} or in turbulence measurements by \citet{venturi_karniadakis_04}.
However, even the best interpolation in terms of the minimum prediction error, understood as the minimum mean square error between the interpolated signal and the true signal, will lead to a significant dynamic error depending on the probability of invalid data points.
The prediction will ideally replace missing samples by something like the expectation of all possible continuations of the signal at the respective time instance.
Then the interpolation itself would be bias-free, however, it will suppress parts of the fluctuations of the true signal.
Finally, statistical properties of the partially interpolated signal deviate from those of the original signal.
For rare and short gaps in the data sequence this may work sufficiently, for more and longer gaps, the error may easily become unacceptable.
Note that this holds for any interpolation scheme, even for those, which perfectly mimic the spectral composition of the signal under observation.

\citet{plantier_etal_12} derive the expected dynamic error particularly for sample-and-hold interpolation, see also \cite{note_on_refinement}.
\footnote{\citet{plantier_etal_12} state that Eq.~(14) (in that publication) would be an approximation. However, it is the explicit and exact formulation for the signal model used by \citet{efficient}, namely the random sampling of a continuous time. The respective formulation for the equidistant sampling used by \citet{plantier_etal_12} is $c_0(\alpha)=1-\frac{2}{\alpha}+\frac{2}{\alpha^2}$}.
For an assumed covariance function of the uninterrupted signal, the covariance function after interpolation of missing data points is predicted, using the statistical properties of the occurrence of these missing data points.
The primary covariance function estimated from the interpolated signal then can be improved by a deconvolution derived from this correspondence.
Under ideal conditions the deconvolution entirely inverts the dynamic error caused by the interpolation.
Therefore, the idea to consider and to invert the influence of the interpolation step, is a promising method to obtain bias-free estimates of the covariance function and the spectrum.
However, the derivation of the mapping matrix depends on specific statistical properties of the data gaps, namely a random occurrence of independent invalid samples. 
If the statistics of the data gaps change, then the procedure needs substantial modification.
Therefore, it is no universal solution.

\citet{sujbert_orosz_16} use no interpolation.
Instead the expected spectrum of the discrete Fourier transform has been derived for the signal with missing samples or data gaps of various statistical characteristics. 
From these expectations, a deconvolution could have been developed to improve the estimation of the spectra, similar to the deconvolution after interpolation by \citet{plantier_etal_12}.
Instead, a procedure has been used avoiding the data gaps by rejecting all values past the first missing sample by means of zero padding.
Depending on the probabilities of the occurrence of the first missing sample, the spectra get different resolution and the superposition of many spectra from individual data blocks becomes smeared.
However, this method is very inefficient in using the information available, since significant amount of valid samples get rejected.
It is limited to small amounts of missing samples anyway, since series of valid samples become too short for increasing amounts of missing samples.

There exists a wide variety of direct spectral estimators optimized for spectral estimation from a limited (typically small) number of unevenly sampled observations of signals \citet{lomb_76,scargle_82,masry_83,ferrazmello_86,foster_96,mathias_etal_04,rivoira_fleury_04,stoica_sandgren_06,babu_stoica_10,LDVprocAC_review}. 
They are widely counted as direct spectral estimators, since the amplitudes of the spectrum are obtained directly at any frequency from the sinusoidal fits.
All these estimators potentially are able to process also signals with missing samples.
However, the spectral composition of equidistantly sampled signals with independently missing samples deviates from random sampling in continuous time, not to mention correlated data gaps, which the above methods cannot handle accurately.
Since Lomb-Scargle's method is widely used as a benchmark, it is included in the comparison below to prove it as biased.

Missing data samples in equidistant data streams have also been investigated broadly by \citet{jones_62a,jones_62b,parzen_63,scheinok_65,bloomfield_70,jones_71,jones_72,ghazal_elhassanein_06,munteanu_etal_16}, including also specific cases of correlated data gaps.
These derivations strictly depend on the specific cases of missing data and are not robust against changes in spectral content of the data gaps.
This also holds for \citet{robinson_77,jones_80,dunsmuir_robinson_81a,dunsmuir_robinson_81b,robinson_85}, who performed parametric estimation. 
This way effective process identification is possible in a limited search space. 
However, bias-free estimation is not possible for unknown or changing spectral content of the data gaps.

In the present article a non-parametric and bias-free method is introduced, which is i) simple to realize, ii) efficient in using the available information and iii) universal for various statistical properties of the data gaps.
It is a combination of three methods a) taken from \citet{jones_71}, except for any spectral or time windowing, 
b) using derivations as in \citet{vogelsang_yang_16} adapted to the two-sided autocovariance function 
including weighting and transferred to the cross-covariance case (see details in \citet{bessel_correction_correlation_arxiv}) and 
c) mentioned only briefly as an appropriate means for spectral estimation in \citet{bartlett_48}.
This combination allows bias-free estimation of the variance of the signal with data gaps, its covariance function and the corresponding power spectrum, independent of the spectral content of the data gaps, including those cases, where the mean value is estimated and subtracted out from the data.
No pre-knowledge about the characteristics of the signal under investigation is needed.
The procedure performs the averaging as it occurs in the definition of the statistical properties, taking into account only valid data samples and masking the invalid ones.
Interpolation of missing portions of the signal is not required.
Neither interpolation of the data gaps nor any kind of reconstruction of the signal are intended. 
Regularization attempts or model-based estimation are not taken into account.

The invalid samples are assumed {\em a priori} known, given by an additional flag (weight) for each data sample indicating whether the sample is valid or invalid.
This way, the samples with invalid values can be excluded from the averaging process.
It is assumed here, that the data gaps (randomly and independently occurring outliers as well as correlated data gaps) are not containing information from the data and that they occur randomly.
While the first assumption is essential for the sufficiency of the data ensemble, no requirements are involved concerning the statistical characteristics or the occurrence of the data gaps.
In extreme cases, the gaps could also occur with a static pattern in consecutive realizations of the signal.
The only requirement to the occurrence of data gaps is that pairs of data points exist at all required time lags for estimating the covariance function.
The programs used here are available as supplementary material to the present article.

Note, that the introduced processing methods are suitable for signals from stationary stochastic processes only and for their statistical analysis. 
They are not suited for signals with time-dependent statistical properties, e.g.\ single pulses. 
For those, missing information cannot be restored by the methods presented here. 
The method is just using that information, which is available.
Note further, that the weighted averages lead to different discrimination of the data ensemble at different lag times.
This results in a correlation matrix, which potentially may violate the non-negative definiteness.
As a consequence, negative values occur in the corresponding power spectral density. 
Since the introduced procedures yield bias-free estimates for both, the covariance function as well as the spectrum, averages over multiple estimates of these functions will converge towards the true functions of the underlying process.
Furthermore, the procedures are shown to be consistent. 
Therefore, the estimated functions will also converge towards the true functions if applied to single but longer data sets, without losing information between block boundaries.
The ultimate solution of course is regularization.
Since this inevitably introduces a bias in both the covariance function and the corresponding spectrum, regularization is not considered in the present article, where bias-free estimation has priority.

\section{Processing method}
\label{sec:gappy}

\begin{table}[!t]
\renewcommand{\arraystretch}{1.05}
\caption{Nomenclature}
\label{tab:nomenclature}
\begin{center}
\begin{tabular}{|l||l|}
\hline
${\vecstyle{x}}$, ${\vecstyle{y}}$& signals (data sequences) without data gaps\\
$x_i$, $y_i$& $i$-th values of the sequences ${\vecstyle{x}}$ and ${\vecstyle{y}}$ resp.\\
&(similar use of indices for other sequences)\\
$t_i$&time instance of the $i$-th sample\\
$\Delta\tau$&fundamental sampling interval\\
$N$&record length (number of values)\\
$N_\rmx$, $N_\rmy$&number of values in the sequences ${\vecstyle{x}}$ and ${\vecstyle{y}}$ resp.\\
${\vecstyle{z}}$& signal with data gaps (valid values from signal ${\vecstyle{x}}$\\
&and individual faulty values)\\
${\vecstyle{w}}$& sequence of weights ($w_i\in\{0;1\}$), indicating\\
&validity of values in sequence ${\vecstyle{z}}$ \\
${\vecstyle{z}}_\rmx$,${\vecstyle{z}}_\rmy$& signals with data gaps from ${\vecstyle{x}}$ and ${\vecstyle{y}}$ resp.\\
${\vecstyle{w}}_\rmx$,${\vecstyle{w}}_\rmy$& weights indicating valid values in ${\vecstyle{z}}_\rmx$ and ${\vecstyle{z}}_\rmx$\\
&resp.\\
$\mu$&true mean value\\
$\mu_\rmx$, $\mu_\rmy$&true mean values of signals ${\vecstyle{x}}$ and ${\vecstyle{y}}$ resp.\\
$\bar z$&estimate of the mean value $\mu$ (obtained from ${\vecstyle{z}}$)\\
$\bar z_\rmx$, $\bar z_\rmy$&estimates of the mean values $\mu_\rmx$ and $\mu_y$ resp.\\
&(obtained from ${\vecstyle{z}}_\rmx$ and ${\vecstyle{z}}_\rmy$)\\
$\sigma^2$&true signal's variance\\
$s^2$&estimate of signal's variance\\
$\hat s^2$&corrected estimate of signal's variance\\
$\sigma_{\bar z}^2$&true variance of the mean estimator $\bar z$\\
${\vecstyle{\gamma}}$&true autocovariance function of signal ${\vecstyle{x}}$\\
${\vecstyle{\gamma}}_\rmxy$&true cross-covariance function of signals ${\vecstyle{x}}$ and ${\vecstyle{y}}$\\
${\vecstyle{C}}$, ${\vecstyle{C}}_\rmxy$&estimates of ${\vecstyle{\gamma}}$ and ${\vecstyle{\gamma}}_\rmxy$ resp.\\
$\hat {\vecstyle{C}}$, $\hat {\vecstyle{C}}_\rmxy$&corrected estimates of ${\vecstyle{\gamma}}$ and ${\vecstyle{\gamma}}_\rmxy$ resp.\\
$\tau_k$&$k$-th lag time\\
$K$&number of lag times considered\\
$K_1$, $K_2$&interval limits of lag times considered\\
${\matstyle{A}}$, ${\matstyle{A}}_\rmxy$&mapping matrix (for the covariance)\\
$a_{kj}$, $a_{\rmxy,kj}$&elements of the mapping matrix\\
$\alpha'$&probability that a sample is valid\\
${\vecstyle{\varepsilon}}$, ${\vecstyle{\varepsilon}}_\rmxy$&systematic error (bias) of the autocovariance and\\
&that of the cross-covariance resp.\\
${\vecstyle{S}}$, ${\vecstyle{S}}_\rmxy$&empirical estimates of the power spectral density\\
&and the cross-power spectral density resp.\\
$f_j$&$j$-th frequency\\
$D$, ${\vecstyle{Z}}$, ${\vecstyle{W}}\!$, ${\vecstyle{G}}$, ${\vecstyle{H}}$&auxiliary values and sequences\\
\hline
\end{tabular}
\end{center}
\end{table}

Assuming a time-limited and equidistantly sampled signal ${\vecstyle{x}}$ yielding a set of $N$ values $x_i$ with $i=0\ldots N-1$.
Sampling instances are the times $t_i=i\Delta\tau$ with the fundamental sampling interval $\Delta\tau$.
The observation ${\vecstyle{z}}$ of this signal literally having also $N$ values $z_i$ is assumed to include missing values.
The corresponding weights $w_i\in\lbrace 0;1\rbrace$ provide information on the validity of the observed values $z_i$.
For $w_i=1$ the value $z_i$ is correct, where $z_i=x_i$.
For $w_i=0$ the value $z_i$ is faulty, where $z_i$ is arbitrary.
Further errors within the measurement process are not investigated.

\subsection{Mean value and variance}

The mean $\mu$ of the true signal ${\vecstyle{x}}$ is defined as
\begin{equation}
\mu=\left\langle {\vecstyle{x}} \right\rangle
\end{equation}
with the expectation $\langle \cdot \rangle$.
Considering appropriate weights $w_i$, the mean value of the measured data set ${\vecstyle{z}}$ with missing data can be obtained based on the valid samples as
\begin{equation}
\bar z=\frac{1}{D}\left(\sum\limits_{i=0}^{N-1} w_i z_i\right)\label{eq:mean}
\end{equation}
with
\begin{equation}
D=\sum\limits_{i=0}^{N-1} w_i.
\end{equation}
If the appropriate weights $w_i$ correspond to the validity of the samples only and do not include other information about the data, then this mean estimator is bias-free.
\begin{equation}
\langle \bar z\rangle=\mu
\end{equation}

The variance of the original signal ${\vecstyle{x}}$ is defined as
\begin{equation}
\sigma^2=\left\langle \left( {\vecstyle{x}}-\mu\right)^2 \right\rangle
\end{equation}
with the true mean value $\mu$ of the signal.
A suitable estimator for the variance using the measured and gappy data set ${\vecstyle{z}}$ considering the weights $w_i$ is
\begin{equation}
s^2=\frac{1}{D}\left[\sum\limits_{i=0}^{N-1} w_i \left(z_i-\bar z\right)^2\right]\label{eq:vari}.
\end{equation}
Here, the true mean value $\mu$ has been replaced by the estimate $\bar z$. 
Even if the mean estimate $\bar z$ is bias-free, the variance estimate $s^2$ is asymptotically bias-free only.
The bias decreases with increasing number of (valid) observations. 
An appropriate correction for short data sets is given in Sec.~\ref{sec:short}.

\subsection{Autocovariance function and spectrum}

The definition of the autocovariance function ${\vecstyle{\gamma}}$ of the true signal ${\vecstyle{x}}$ is
\begin{equation}
\gamma_k=\left\langle \left( x_i - \mu \right) \left( x_{i+k}-\mu \right)\right\rangle
\end{equation}
with the expectation $\langle \cdot \rangle$ and the true mean value $\mu$ of signal.
The autocovariance function is also equidistantly sampled with the sampling interval $\Delta\tau$.
%
%
%
Considering the weights $w_i$, the empirical autocovariance function ${\vecstyle{C}}$ can be obtained directly from the measured and gappy data set ${\vecstyle{z}}$ through
\begin{equation}
C_k=\frac{Z_k}{W_k}\label{eq:autocovariance}
\end{equation}
for all lag times $\tau_k=k\Delta\tau$ with
\begin{eqnarray}
Z_k&=&\sum\limits_{i=I_1}^{I_2} w_i w_{i+k} (z_i-\bar z) (z_{i+k}-\bar z)\\
W_k&=&\sum\limits_{i=I_1}^{I_2} w_i w_{i+k}
\end{eqnarray}
with $I_1=\max(0,-k)$, $I_2=\min(N,N-k)-1$ and with the estimated mean value $\bar z$ as above.
To accelerate the computing, the values $Z_k$ and $W_k$ can also be obtained as the $k$-th values of the records
\begin{eqnarray}
{\vecstyle{Z}}&=&\opstyle{IFFT}\left\lbrace \abs{\opstyle{FFT}\left\lbrace \underline{\vecstyle{w}}\cdot\left(\underline{\vecstyle{z}-\bar z}\right) \right\rbrace }^2\right\rbrace\\
{\vecstyle{W}}\!&=&\opstyle{IFFT}\left\lbrace \abs{\opstyle{FFT}\left\lbrace \underline{\vecstyle{w}}\right\rbrace }^2\right\rbrace
\end{eqnarray}
with the element-wise product $\cdot$, where the fast discrete Fourier transform (FFT) and its inverse (IFFT) are utilized. 
The records $\underline{\vecstyle{z}-\bar z}$ and $\underline{\vecstyle{w}}$ are the mean-subtracted, measured and gappy signal ${\vecstyle{z}}-\bar z$ and the sequence of weights ${\vecstyle{w}}$, both expanded by zero padding, concatenating $N$ zeros to each of the two sequences.
This way the values $Z_k$ and $W_k$ are obtained for all $k=-N\ldots N-1$ in one step.

Following Wiener-Khinchin's theorem, see \citet{khinchin_34} --- the transcription of the name from Cyrillic to Latin letters is ambiguous ---
the power spectral density can be obtained from the autocovariance function through the discrete Fourier transform.
Unfortunately, the estimation of the power spectral density from the autocovariance function obtained from a single data set has an unacceptable high estimator variance. 
A common means to reduce the variance is a subdivision of the data set into shorter blocks.
The average of the power spectral densities of the data blocks then has a significantly smaller variance.
This method is known as Bartlett's method, see \citet{bartlett_48,bartlett_50}.
A disadvantage of Bartlett's method is that correlations between the samples at the end of one block and the beginning of the next are not counted.
Furthermore, the wrap-around error may be increased if the assumption is made that the signal respectively the block is periodic. 
For too short blocks this may lead to significant deviations.
In contrast, for longer blocks the reduction of the estimator variance becomes less effective.
With Welch's method, see \citet{welch_67}, where the statistical functions from overlapping blocks get averaged, correlations between block boundaries are counted. 
However, this also partially generates redundancy. 
This fact is taken into account by applying windowing functions to the data blocks prior to their statistical analysis.
Unfortunately, the additional modulation of the data introduces an additional bias to the estimates of the statistical functions.
%
%
However, this is computationally costly.

A powerful alternative without the necessity of block subdivision is the reduction of the spectral resolution in a post-processing step. 
Reducing the spectral resolution corresponds to a shorter support of the covariance function. 
For a random process with arbitrarily long but finite memory the autocovariance function is zero at longer lag times.
In this case the autocovariance function can be shortened to the extent of the longest lasting correlation without losing information.
Assuming that the empirical autocovariance function ${\vecstyle{C}}$, computed for $K$ lag times with $K\ll 2N$ (The factor 2 comes from previous zero padding of the signal.) is negligible outside the interval $k=-\lfloor K/2\rfloor \ldots \lfloor (K-1)/2\rfloor$ with $\lfloor i\rfloor$ being the largest integer smaller or equal to $i$.
Then the corresponding empirical power spectral density ${\vecstyle{S}}$ can be obtained from the shorter autocovariance function ${\vecstyle{C}}$ via the discrete (fast) Fourier transform (FFT)
\begin{equation}
{\vecstyle{S}}=\Delta\tau\cdot\opstyle{FFT}\lbrace {\vecstyle{C}}\rbrace
\label{eq:psd}
\end{equation}
with
\begin{equation}
S_j=\Delta\tau\cdot\sum\limits_{k=-\lfloor K/2\rfloor}^{\lfloor (K-1)/2\rfloor} C_k \exp(-2\pi\iu f_j\tau_k)
\end{equation}
for the frequencies $f_j=j/(K\Delta \tau)$ with $j=-\lfloor K/2\rfloor\ldots \lfloor (K-1)/2\rfloor$ and with the imaginary unit $\iu$.
Due to the shorter support of the autocovariance function, the spectral resolution is reduced accordingly, leading also to a significantly lower estimator variance of the spectrum without introducing new errors from too short a block subdivision or any modulation of the signal's amplitude.
The advantage of this method compared to usual block subdivision is that the correlations of all samples up to a certain maximum lag time are considered from the entire data set or, at least, from sufficiently long blocks.
Further suppression of the wrap-around error e.g.\ via the application of a window function is not needed. 
This way, one also avoids the spectrum becoming smeared by additional modulation of the signal.
This method is identical to \citet{blackman_tukey_58_part_I,blackman_tukey_58_part_II} used with a rectangular window applied to the covariance function estimated.
Furthermore, this method has been previously investigated by and briefly mentioned in \citet{bartlett_48}, yielding results comparable to block averaging.
Since longer data records need to be Fourier transformed, truncation of primarily long covariance estimates is computationally more expensive.
However, it is superior in efficiently using the information available.

Since the discrete Fourier transform is linear, the above transform ensures that a bias-free estimate of the covariance function also leads to a bias-free estimate of the spectrum, except for possible aliasing errors.
Note, that this is valid only under the condition that all correlations of the investigated process are within the support of the estimated covariance function.
For longer lasting correlations and too short support of the covariance estimate, the individual values of the estimate still may be bias-free.
However, the spectrum gets smeared if correlations outside the investigated support are missing.
Remaining aliasing errors are not further investigated, since they occur with too a low temporal resolution of the fundamental sampling process. 
Instead, the estimation from time series with invalid samples is called bias-free, if the estimates have an expectation identical to the equidistantly sampled process without missing data. 
This includes appropriate reproduction of possible aliasing errors.

\subsection{Cross-covariance function and spectrum}

For two signals ${\vecstyle{x}}$ and ${\vecstyle{y}}$, the cross-covariance function ${\vecstyle{\gamma}}_\rmxy$ is defined as
\begin{equation}
\gamma_{\rmxy,k}=\left\langle \left( x_i - \mu_\rmx \right) \left( y_{i+k}-\mu_\rmy \right)\right\rangle
\end{equation}
with the expectation $\langle \cdot \rangle$ and the true mean values $\mu_\rmx$ of signal ${\vecstyle{x}}$ and $\mu_\rmy$ of signal ${\vecstyle{y}}$ respectively.
From the two data sets $z_{\rmx,i}$ with $i=0\ldots N_\rmx-1$ and $z_{\rmy,i}$ with $i=0\ldots N_\rmy-1$ including their appropriate weights $w_{\rmx,i}$ and $w_{\rmy,i}$, the empirical cross-covariance function ${\vecstyle{C}}_\rmxy$ can be estimated directly as
\begin{equation}
C_{\rmxy,k}=\frac{Z_{\rmxy,k}}{W_{\rmxy,k}}
\label{eq:crosscovariance}
\end{equation}
for all lag times $\tau_k=k\Delta\tau$ with
\begin{eqnarray}
Z_{\rmxy,k}&=&\sum\limits_{i=I_1}^{I_2} w_{\rmx,i} w_{\rmy,i+k} (z_{\rmx,i}-\bar z_\rmx) (z_{\rmy,i+k}-\bar z_\rmy)\\
W_{\rmxy,k}&=&\sum\limits_{i=I_1}^{I_2} w_{\rmx,i} w_{\rmy,i+k}
\end{eqnarray}
with $I_1=\max(0,-k)$, $I_2=\min(N_\rmx,N_\rmy-k)-1$ and with the estimated mean values $\bar z_\rmx$ and $\bar z_\rmy$ obtained from the data sets $z_\rmx$ and $z_\rmy$ and the appropriate sequences of weights $w_\rmx$ and $w_\rmy$ using the weighted mean estimator in Eq.~(\ref{eq:mean}).
Using zero padding and the (inverse) discrete (fast) Fourier transform (FFT resp.\ IFFT), the values $Z_{\rmxy,k}$ and $W_{\rmxy, k}$ can also be obtained as the $k$-th values of the records
\begin{eqnarray}
{\vecstyle{Z}}_\rmxy&=&\opstyle{IFFT}\left\lbrace \opstyle{FFT}\left\lbrace \underline{\vecstyle{w}_\rmx}\cdot\left(\underline{\vecstyle{z}_\rmx-\bar z_\rmx}\right) \right\rbrace ^\ast  \cdot\opstyle{FFT}\left\lbrace \underline{\vecstyle{w}_\rmy}\cdot\left(\underline{\vecstyle{z}_\rmy-\bar z_\rmy}\right) \right\rbrace \right\rbrace\\
{\vecstyle{W}\!}_\rmxy&=&\opstyle{IFFT}\left\lbrace \opstyle{FFT}\left\lbrace \underline{\vecstyle{w}_\rmx} \right\rbrace ^\ast \cdot\opstyle{FFT}\left\lbrace \underline{\vecstyle{w}_\rmy} \right\rbrace \right\rbrace
\end{eqnarray}%
with the element-wise product $\cdot$ and the conjugate complex $^\ast$.
The underlined terms are the sequences after zero padding.
These are the sequences ${\vecstyle{z}}_\rmx-\bar z_\rmx$ and ${\vecstyle{w}}_\rmx$ extended by $N_\rmy$ concatenated zeros and ${\vecstyle{z}}_\rmy-\bar z_\rmy$ and ${\vecstyle{w}}_\rmy$ extended by $N_\rmx$ concatenated zeros, yielding sequences of identical length.

The empirical cross-covariance function ${\vecstyle{C}}_\rmxy$, computed for $K$ lag times $k=K_1\ldots K_2$, can finally be transformed into the corresponding empirical power spectral density ${\vecstyle{S}}_\rmxy$ by the discrete (fast) Fourier transform (FFT)
\begin{equation}
{\vecstyle{S}}_\rmxy=\Delta\tau\cdot\opstyle{FFT}\lbrace {\vecstyle{C}}_\rmxy\rbrace
\label{eq:xpsd}
\end{equation}
with
\begin{equation}
S_{\rmxy,j}=\Delta\tau\cdot\sum\limits_{k=K_1}^{K_2} C_{\rmxy,k}\exp(-2\pi\iu f_j\tau_k)
\end{equation}
for the frequencies $f_j=j/(K\Delta \tau)$ with $j=-\lfloor K/2\rfloor\ldots \lfloor (K-1)/2\rfloor$ and with the imaginary unit $\iu$.
Requirements according the choice of the investigated interval of the covariance function are identical to those for the autocovariance case above, namely no correlations outside the investigated interval.

\section{Correction for short data sets (Bessel's correction)}
\label{sec:short}

The estimators above are asymptotically bias-free only.
For short data records, the amount of valid data may become insufficient and the bias may become significant.
For the estimators introduced above an appropriate correction is given below.
However, other estimators could also benefit from this correction in the case of short data records.
Furthermore, the following corrections are applied here to gappy data sets only with weights $w_i\in\lbrace 0;1\rbrace$.
However, the procedures are suitable also for other choices of weights including non-binary values.

\subsection{Variance estimate}

Let the mean estimator of Eq.~(\ref{eq:mean}) have the estimator variance $\sigma_{\bar z}^2$.
Since the variance of a sum of correlated variables is the sum of all pair-wise covariances, the variance of the sum in the numerator of Eq.~(\ref{eq:mean}) is $\sum_{i=0}^{N-1}\sum_{j=0}^{N-1} w_i w_j \gamma_{j-i}$
involving the unknown true autocovariance function $\gamma_k={\vecstyle{\gamma}}(\tau_k)$ with $\tau_k=k\Delta\tau$.
The denominator in Eq.~(\ref{eq:mean}) applies twice, to each factor of the pair-wise covariances, yielding finally the variance of the mean estimator
\begin{equation}
\sigma_{\bar z}^2=\frac{1}{D^2}\left(\sum\limits_{i=0}^{N-1}\sum\limits_{j=0}^{N-1} w_i w_j \gamma_{j-i}\right).
\label{eq:meanvari}
\end{equation}

The expectation of the variance estimation as in Eq.~(\ref{eq:vari}) with the estimated mean subtracted from the data is (for the derivation see App.~\ref{sec:appexpvari})
\begin{equation}
\langle s^2\rangle =\sigma^2-\frac{1}{D^2}\left(\sum\limits_{i=0}^{N-1}\sum\limits_{j=0}^{N-1}w_i w_j\gamma_{j-i}\right)
\label{eq:esq}
\end{equation}
with the true variance $\sigma^2$ of the data and again with the true autocovariance function ${\vecstyle{\gamma}}$.
The deviation from the correct variance is exactly the variance of the mean estimator $\sigma_{\bar z}^2$.

If the variance of the mean estimation is known beforehand, then a bias-free estimate of the data variance is\footnote{If all samples are independent and all weights are one, this leads to the bias-free estimator
$\hat s^2=\frac{1}{N-1}\cdot\sum_{i=0}^{N-1} \left(z_i-\bar z\right)^2$, where the division by $N-1$ instead of $N$ is widely known as Bessel's correction for the variance estimate for independent samples.
Similar corrections can be made to estimates of the autocovariance function or the cross-covariance function derived from two different data sets. 
Unfortunately, this requires considering that the data samples are correlated --- why one would otherwise calculate the covariance function?}
\begin{equation}
\hat s^2=s^2+\sigma_{\bar z}^2.
\label{eq:varicorrect}
\end{equation}
For unknown a variance of the mean estimation, first the autocovariance estimate must be corrected.
Therefore, the recipe to obtain a bias-free estimate of the variance is given at the end of Sec.~\ref{sec:autobessel}, first introducing the correction for the autocovariance estimate.

\subsection{Autocovariance function and spectrum}
\label{sec:autobessel}

\citet{vogelsang_yang_16} derived a procedure of an exact correction for short data sets for the autocovariance estimation.
This procedure has been further developed in \citet{bessel_correction_correlation_arxiv}, for the two-sided autocovariance function, 
correcting the autocovariance estimate normalized with the number of pairs of values instead of a constant, 
including sample weights and also extended to the cross-covariance case. 

The autocovariance estimator in Eq.~(\ref{eq:autocovariance}) has the expectation
\begin{equation}
\langle C_k\rangle =\gamma_k+\varepsilon_k
\label{eq:eck}
\end{equation}
with the true autocovariance function ${\vecstyle{\gamma}}$ and the bias (for the derivation see App.~\ref{sec:appcov})
\begin{eqnarray}
\varepsilon_k&=&\frac{1}{D^2}\left(\sum\limits_{i=0}^{N-1}\sum\limits_{j=0}^{N-1}w_i w_j \gamma_{j-i}\right)\nonumber\\
&&-\frac{1}{D W_k}\left[\sum\limits_{i=I_1}^{I_2}\sum\limits_{j=0}^{N-1}w_i w_{i+k} w_j (\gamma_{j-i}+\gamma_{i+k-j})\right]\quad
\label{eq:eek}
\end{eqnarray}
with $I_1=\max(0,-k)$, $I_2=\min(N,N-k)-1$.
The value is constant for uncorrelated data, otherwise it varies with $k$.
The first term again is the variance $\sigma_{\bar z}^2$ of the mean estimator.
Since the true covariance function ${\vecstyle{\gamma}}$ is unknown in real measurements, the prediction cannot be made directly. 
However, the relation between the true covariance function and its estimate is linear. 
Therefore, one can built a matrix ${\matstyle{A}}$, mapping a hypothetical covariance function ${\vecstyle{\gamma}}$ onto the expectation of the estimated one $\langle {\vecstyle{C}}\rangle$.
\begin{equation}
\langle {\vecstyle{C}}\rangle ={\matstyle{A}} {\vecstyle{\gamma}},
\end{equation}
If the matrix ${\matstyle{A}}$ has the elements $a_{kj}$ then the prediction of the estimated covariance at lag time $\tau_k$ is
\begin{equation}
\langle C_k\rangle =\sum_{j=K_1}^{K_2}a_{kj} \gamma_j.
\end{equation}
The range $K_1\ldots K_2$ of covariances considered should again include the full range of occurring correlations, such that all true covariances outside this interval can be neglected.
If one sorts the summands in Eq.~(\ref{eq:eck}) and Eq.~(\ref{eq:eek}) for an increasing index of the true autocovariance ${\vecstyle{\gamma}}$, one obtains the elements of the matrix
\begin{equation}
a_{kj}=\delta_{k-j}+\frac{W_j}{D^2}-\frac{G_{kj}+H_{kj}}{D W_k}
\end{equation}
with
\begin{equation}
\delta_{k-j}=\left\lbrace \begin{array}{ll}1&\mbox{for $k-j=0$}\\0&\mbox{otherwise}\end{array}\right.
\end{equation}%
and with $W_j$ and $W_k$ being the $j$-th and the $k$-th values of the record ${\vecstyle{W}}\!$ above and
\begin{eqnarray}
G_{kj}&=& \sum\limits_{i=\max(0,-j,-k)}^{\min(N,N-j,N-k)-1} \hspace{-2.5mm} w_i w_{i+j} w_{i+k}\\
H_{kj}&=& \sum\limits_{i=\max(0,-j,k-j)}^{\min(N,N-j,N+k-j)-1} \hspace{-2.5mm} w_i w_{i+j} w_{i+j-k} .
\end{eqnarray}
Alternatively, the values $G_{kj}$ and $H_{kj}$ can be obtained as the $j$-th values of the records
\begin{eqnarray}
{\vecstyle{G}}_k&=&\opstyle{IFFT}\Big\lbrace \opstyle{FFT}\left\lbrace \underline{\vecstyle{w}}\cdot\underline{\vecstyle{w}}_{\,+k} \right\rbrace ^\ast \cdot\opstyle{FFT}\left\lbrace \underline{\vecstyle{w}} \right\rbrace \Big\rbrace \\
{\vecstyle{H}}_k&=&\opstyle{IFFT}\Big\lbrace \opstyle{FFT}\left\lbrace \underline{\vecstyle{w}} \right\rbrace ^\ast \cdot\opstyle{FFT}\left\lbrace \underline{\vecstyle{w}}\cdot\underline{\vecstyle{w}}_{\,-k}\right\rbrace \Big\rbrace
\end{eqnarray}
with the element-wise product $\cdot$ and the conjugate complex $^\ast$, involving again the (fast) discrete Fourier transform (FFT) and its inverse (IFFT).
The record $\underline{\vecstyle{w}}$ again is the sequence of weights ${\vecstyle{w}}$, expanded by zero padding, concatenating $N$ zeros.
The records $\underline{\vecstyle{w}}_{\,+k}$ and $\underline{\vecstyle{w}}_{\,-k}$ are the sequence of weights, expanded by zero padding and then shifted by $+k$ or $-k$ time steps respectively. 
While $\underline{\vecstyle{w}}$ at index $i$ has the weight $w_i$, $\underline{\vecstyle{w}}_{\,+k}$ has the weight $w_{i+k}$ at index $i$ and $\underline{\vecstyle{w}}_{\,-k}$ has the weight $w_{i-k}$.

The inverse ${\matstyle{A}}^{-1}$ of the matrix ${\matstyle{A}}$ applied to the primary estimate ${\vecstyle{C}}$ yields an improved, bias-free estimate $\hat {\vecstyle{C}}$ of the covariance
\begin{equation}
\hat {\vecstyle{C}}={\matstyle{A}}^{-1} {\vecstyle{C}}.
\end{equation}

For given $N$ samples $z_i$, the covariance function after zero padding has $2N-1$ non-zero values $C_k$ in the range $-(N-1)\ldots N-1$. 
Unfortunately, the appropriate matrix ${\matstyle{A}}$ then is singular and the inverse does not exist for this case.
The inverse can be calculated only, if the covariance function is limited to the range $K_1\ldots K_2$ with $-(N-1)<K_1\le K_2<N-1$. 
The improved covariance estimate then is bias-free, as long as the true covariance of the original signal is zero outside the reduced interval of lag times $\tau_{K_1}\ldots \tau_{K_2}$. 
This coincides with the requirement that the interval of investigated lag times is larger than the longest correlation lasts and the observation interval of the signal is longer than the largest lag time investigated.

The improved estimate $\hat {\vecstyle{C}}$ of the covariance function then can be used to derive the variance of the mean estimator $\sigma_{\bar z}^2$ following Eq.~(\ref{eq:meanvari}), where the true covariance ${\vecstyle{\gamma}}$ is replaced by the improved, bias-free estimate $\hat {\vecstyle{C}}$, and finally to improve the estimation $\hat s^2$ of the variance following Eq.~(\ref{eq:varicorrect}).
Note, that for the weights being only zero or one, the relations $C_0=s^2$ and $\hat C_0=\hat s^2$ hold.
For other weights, $C_0$ may differ from $s^2$ as well as $\hat C_0$ from $\hat s^2$.

Finally, the corresponding spectrum is obtained again via the discrete Fourier transform using Eq.~(\ref{eq:psd}).

\subsection{Cross-covariance function and spectrum}

The cross-covariance estimator in Eq.~(\ref{eq:crosscovariance}) has the expectation
\begin{equation}
\langle C_{\rmxy,k}\rangle =\gamma_{\rmxy,k}+\varepsilon_{\rmxy,k}
\label{eq:eckx}
\end{equation}
with the true cross-covariance function ${\vecstyle{\gamma}}_{\rmxy}$ and the bias (for the derivation see App.~\ref{sec:appcovx})
\begin{eqnarray}
\varepsilon_{\rmxy,k}&=&\frac{1}{D_\rmx D_\rmy}\left(\sum\limits_{i=0}^{N_\rmx-1}\sum\limits_{j=0}^{N_\rmy-1} w_{\rmx,i} w_{\rmy,j} \gamma_{\rmxy,j-i}\right)\nonumber\\
&&-\frac{1}{D_\rmy W_{\rmxy,k}}\left[\sum\limits_{i=I_1}^{I_2}\sum\limits_{j=0}^{N_\rmy-1} w_{\rmx,i} w_{\rmy,i+k} w_{\rmy,j} \gamma_{\rmxy,j-i}\right]\nonumber\\
&&-\frac{1}{D_\rmx W_{\rmxy,k}}\left[\sum\limits_{i=I_1}^{I_2}\sum\limits_{j=0}^{N_\rmx-1} w_{\rmx,i} w_{\rmy,i+k} w_{\rmx,j} \gamma_{\rmxy,i+k-j}\right]\quad
\label{eq:eekx}
\end{eqnarray}
with $I_1=\max(0,-k)$, $I_2=\min(N_\rmx,N_\rmy-k)-1$.
The value is constant for uncorrelated data and only if the weights are all identical for the two data sets, otherwise it varies with $k$.
Here
\begin{eqnarray}
D_\rmx&=&\sum\limits_{i=0}^{N_\rmx-1} w_{\rmx,i}\\
D_\rmy&=&\sum\limits_{i=0}^{N_\rmx-1} w_{\rmy,i}
\end{eqnarray}
and $W_{\rmxy,k}$ as above are used for short notation.
The matrix ${\matstyle{A}}_\rmxy$, mapping a hypothetical covariance function ${\vecstyle{\gamma}}_\rmxy$ onto the expectation of the estimated one ${\vecstyle{C}}_\rmxy$ via
\begin{equation}
\langle {\vecstyle{C}}_\rmxy\rangle ={\matstyle{A}}_\rmxy {\vecstyle{\gamma}}_\rmxy
\end{equation}
can be used to predict the estimated covariance at lag time $\tau_k$ as
\begin{equation}
\langle C_{\rmxy,k}\rangle =\sum_{j=K_1}^{K_2}a_{\rmxy,kj} \gamma_{\rmxy,j}
\end{equation}
with the elements $a_{\rmxy,kj}$ of the matrix ${\matstyle{A}}_\rmxy$.
The range $K_1\ldots K_2$ of covariances considered should again include the full range of occurring correlations, such that all true covariance outside this interval can be neglected.

Sorting the summands in Eq.~(\ref{eq:eckx}) and Eq.~(\ref{eq:eekx}) for an increasing index of the true cross-covariance ${\vecstyle{\gamma}}_\rmxy$ yields
\begin{equation}
a_{\rmxy,kj}=\delta_{k-j}+\frac{W_{\rmxy,j}}{D_\rmx D_\rmy}
-\frac{G_{\rmxy,kj}}{D_\rmy W_{\rmxy,k}}
-\frac{H_{\rmxy,kj}}{D_\rmx W_{\rmxy,k}}
\end{equation}
with $\delta_{k-j}$ as above and with $W_{\rmxy,j}$ and $W_{\rmxy,k}$ being the $j$-th and the $k$-th value of the record ${\vecstyle{W}}\!_\rmxy$ above and
\begin{eqnarray}
G_{\rmxy,kj}&=& \sum\limits_{i=\max(0,-j,-k)}^{\min(N_\rmx,N_\rmy-j,N_\rmy-k)-1} \hspace{-7.5mm} w_{\rmx,i} w_{\rmy,i+j} w_{\rmy,i+k}\\
H_{\rmxy,kj}&=& \sum\limits_{i=\max(0,-j,k-j)}^{\min(N_\rmx,N_\rmy-j,N_\rmx+k-j)-1} \hspace{-7.5mm} w_{\rmx,i} w_{\rmy,i+j} w_{\rmx,i+j-k}
\end{eqnarray}
Alternatively, $G_{\rmxy,kj}$ and $H_{\rmxy,kj}$ can be obtained as the $j$-th values of the records
\begin{eqnarray}
{\vecstyle{G}}_{\rmxy,k}&=&\opstyle{IFFT}\left\lbrace \opstyle{FFT}\left\lbrace \underline{\vecstyle{w}_\rmx}\cdot\underline{\vecstyle{w}_{\rmy}}{}_{\,,+k} \right\rbrace ^\ast \cdot\opstyle{FFT}\left\lbrace \underline{\vecstyle{w}_\rmy} \right\rbrace \right\rbrace\qquad \\
{\vecstyle{H}}_{\rmxy,k}&=&\opstyle{IFFT}\left\lbrace \opstyle{FFT}\left\lbrace \underline{\vecstyle{w}_\rmx} \right\rbrace ^\ast \cdot\opstyle{FFT}\left\lbrace \underline{\vecstyle{w}_\rmy}\cdot\underline{\vecstyle{w}_{\rmx}}{}_{\,,-k} \right\rbrace \right\rbrace ,\qquad
\end{eqnarray}
involving again the (fast) discrete Fourier transform (FFT) and its inverse (IFFT) and where again, $\underline{\vecstyle{w}_\rmx}$ and $\underline{\vecstyle{w}_\rmy}$ are the weights ${\vecstyle{w}}_\rmx$ and ${\vecstyle{w}}_\rmy$ after zero padding and $\underline{\vecstyle{w}_{\rmx}}{}_{\,,-k}$ and $\underline{\vecstyle{w}_{\rmy}}{}_{\,,+k}$ are additionally shifted by $-k$ or $+k$ time steps respectively.
While $\underline{\vecstyle{w}_\rmx}$ and $\underline{\vecstyle{w}_\rmy}$ at index $i$ have the weights $w_{\rmx,i}$ and $w_{\rmy,i}$, $\underline{\vecstyle{w}_{\rmx}}{}_{\,,-k}$ has the weight $w_{\rmx,i-k}$ at index $i$ and $\underline{\vecstyle{w}_{\rmy}}{}_{\,,+k}$ has the weight $w_{\rmy,i+k}$.

The inverse ${\matstyle{A}}_\rmxy^{-1}$ of the matrix ${\matstyle{A}}_\rmxy$ applied to the estimate ${\vecstyle{C}}_\rmxy$ yields an improved, bias-free estimate $\hat {\vecstyle{C}}_\rmxy$ of the cross-covariance
\begin{equation}
\hat {\vecstyle{C}}_\rmxy={\matstyle{A}}_\rmxy^{-1} {\vecstyle{C}}_\rmxy.
\end{equation}

For given $N_\rmx$ samples $z_{\rmx,i}$ and $N_\rmy$ samples $z_{\rmy,i}$, the covariance function after zero padding has $N_\rmx+N_\rmy-1$ non-zero values $c_{\rmxy,k}$ in the range $-(N_\rmx-1)\ldots N_\rmy-1$. 
Unfortunately, the appropriate matrix ${\matstyle{A}}_\rmxy$ then is singular and the inverse does not exist for this case.
The inverse can be calculated only, if the covariance function is limited to the range $K_1\ldots K_2$ with $-(N_\rmx-1)<K_1\le K_2<N_\rmy-1$. 
The improved covariance estimate then is bias-free, as long as the true covariance of the original signal is zero outside the reduced interval of lag times $\tau_{K_1}\ldots \tau_{K_2}$. 
This coincides with the requirement that the interval of investigated lag times is larger than the longest correlation lasts and the observation interval of the signal is longer than the largest lag time investigated.

Finally, the corresponding spectrum is obtained again via the discrete Fourier transform using Eq.~(\ref{eq:xpsd}).

\section{Simulation}

To demonstrate the ability of the estimation routines to derive consistent estimates of the covariance functions and the power spectral densities, a moving-average stochastic process is generated from noise and analyzed in Monte-Carlo runs.
This includes both, bias-free estimation and estimator variance decreasing with increasing amount of information within the signal.
Both aspects can be measured by means of the root mean square error (RMS), shown in Sec.~\ref{sec:RMS}.
However, since systematic and random errors get mixed, the plots of the RMS error alone are not specific enough to identify individual features of the various estimators. 
%
%
For systematic errors, which are small in absolute magnitude but significant in comparison to a small true value, or for small random errors around a biased value, the characteristic dynamics of the investigated process may get affected significantly, e.g.\ low-pass filtered. 
Characteristic dynamic features of the process under observation may get concealed this way.
%
Therefore, the next section investigates systematic errors first, before RMS errors are investigated separately.
The empirical mean estimates are shown in direct comparison to the expected values of the simulation.
This way, the main features of the estimators get more obvious, especially dynamic errors.

\subsection{Systematic errors}

\begin{figure}
\centering{%
\includegraphics[scale=0.79]{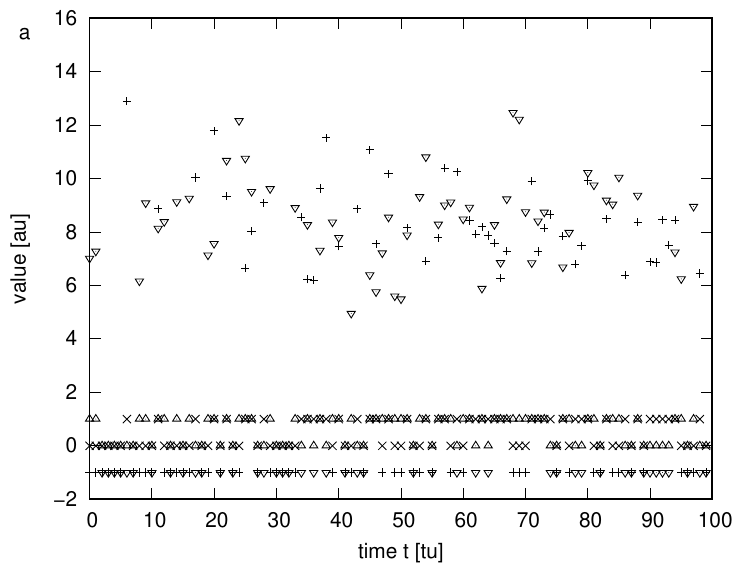}%
\hfill
\includegraphics[scale=0.79]{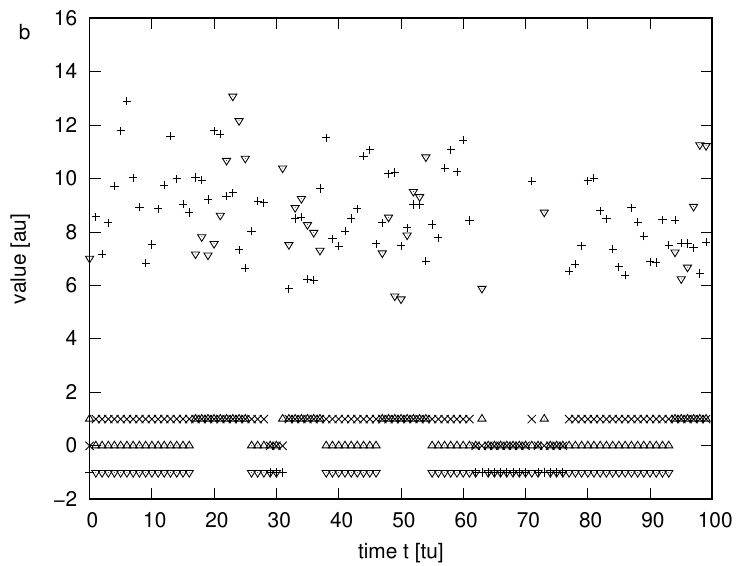}%
}
\caption{Single realizations of the signals with (a) independent outliers and (b) with long gaps ($\unit{au}$ - amplitude unit, $\unit{tu}$ - time unit, 
$+$ data $z_\rmx$, $\times$ weights $w_\rmx$, $\bigtriangledown$ data $z_\rmy$, $\bigtriangleup$ weights $w_\rmy$)}
\label{fig:simmi}
\end{figure}

The simulation uses an autoregressive linear stochastic process of order 100, yielding an artificial spectrum with an exponentially increasing slope and with an artificial dip in the observed frequency range.
The simulated autoregressive process has an infinite impulse response and, therefore, clearly mismatches the assumption of having no correlations outside the observation interval.
This mismatch has been chosen on purpose, to demonstrate that the introduced estimation methods, especially the correction for short data sets, are robust against cases, where processes with remaining long lasting correlations are observed in smaller time windows.
The parameters of the process and the values of the correlation functions and the spectrum are provided as supplementary data with the present article.
Each run of the simulation process generates two signals with such spectral characteristic with a total length of $\unit[100]{tu}$ (time units).
The signals have a mean value of $\unit[8]{au}$ (amplitude units), a variance of $\unit[4]{au^2}$, a cross-covariance of $\unit[3]{au^2}$ and a time delay between the corresponding signals of $\unit[10]{tu}$.
Then for each sample of the two primary signals a weight (1 for a valid sample and 0 for an invalid one) is chosen from a random process to mimic the data gaps.
The first simulation generates individual samples marked as invalid, independently from each other, with a probability of $\unit[50]{\%}$.
The second simulation generates series of invalid samples, where the state of validity changes with a probability of $\unit[10]{\%}$ at each time step.
This procedure also yields $\unit[50]{\%}$ invalid samples on average, where the length of valid data or that of invalid data has an exponential distribution with a mean of ten samples and the sequence of weights gets correlated.
For invalid data points, the samples in the data series are set to the constant value of $\unit[-1]{au}$ for visualization purposes and to cross check that the results get not affected by these faulty values.
In Fig.~\ref{fig:simmi} individual realizations of the signals and the weights are shown.
While in Fig.~\ref{fig:simmi}a only individual samples are marked independently as invalid, in Fig.~\ref{fig:simmi}b longer sequences of invalid samples can be seen.
However, in both cases $\unit[50]{\%}$ of the samples get marked as invalid on average.

The data sets have been processed to auto- and cross-covariance functions for the range of lag times between $\unit[-25]{tu}$ and $\unit[24]{tu}$ ($K=50$) respectively $\unit[-20]{tu}$ and $\unit[29]{tu}$ ($K_1=-20$, $K_2=29$).
The corresponding power spectral densities have been obtained in the range between $\unit[-0.5]{tu^{-1}}$ and $\unit[0.48]{tu^{-1}}$.

\begin{figure}
\centering{%
\includegraphics[scale=0.79]{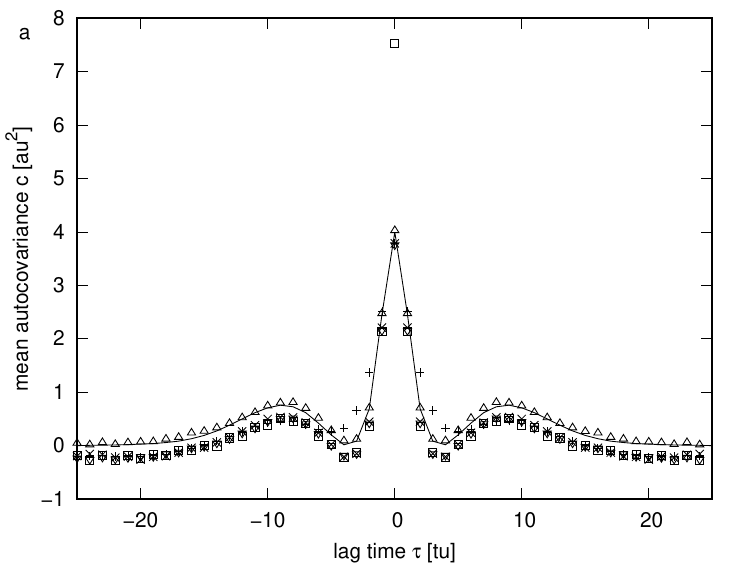}%
\hfill
\includegraphics[scale=0.79]{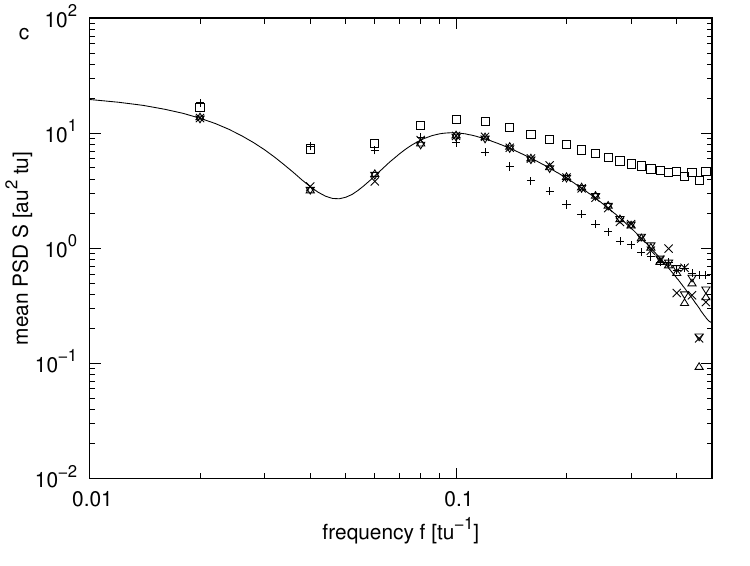}%
}\\
\centering{%
\includegraphics[scale=0.79]{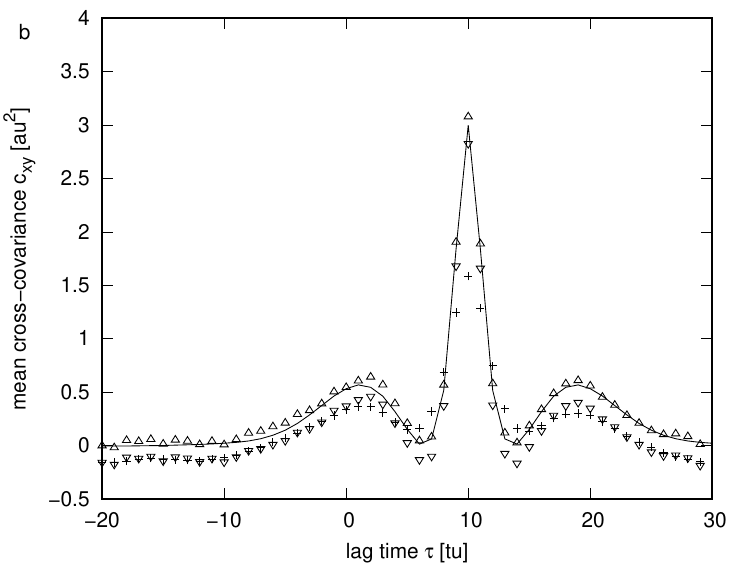}%
\hfill
\includegraphics[scale=0.79]{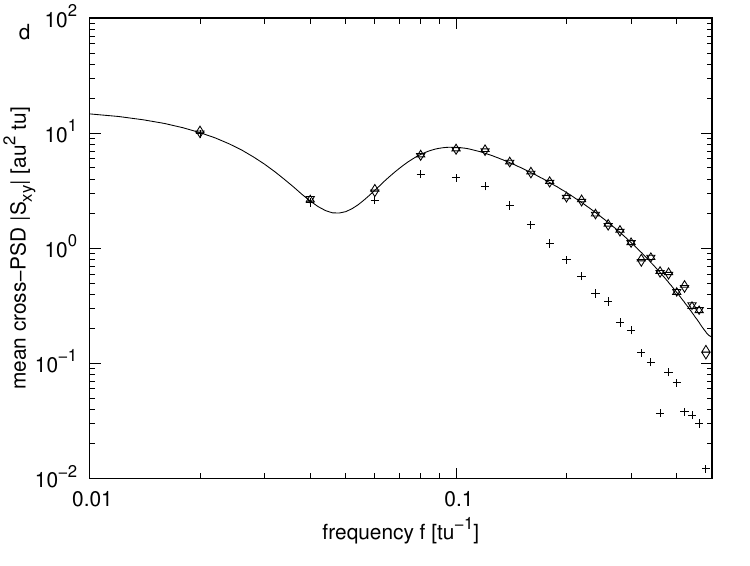}%
}
\caption{Empirical mean from averages over 1\,000 realizations with independent outliers and 100 samples each: 
(a) autocovariance function, (b) cross-covariance function, (c) power spectral density (autospectrum), (d) magnitude of the cross power spectral density (cross-spectrum,
$\unit{au}$ - amplitude unit, $\unit{tu}$ - time unit,
$+$ sample-and-hold interpolation w/o deconvolution, $\times$ sample-and-hold interpolation w deconvolution, $\square$ Lomb-Scargle's method, $\bigtriangledown$ only valid points w/o correction for short data sets, $\bigtriangleup$ only valid points w correction for short data sets, --- simulation).}
\label{fig:simmrand}
\end{figure}

Fig.~\ref{fig:simmrand} shows the empirical mean values of the estimated autocovariance function, that of the cross-covariance function, the auto-power spectral density and the cross-power spectral density obtained from 1\,000 generated data sets for the case of independent outliers.
To simplify the identification of dynamic errors only one half of the spectra are shown in log-log plots.
The introduced estimation routines have been used for the analysis, including the correction for short data sets.
For comparison, the results of Lomb-Scargle's method are also shown as well as those of the sample-and-hold interpolation.
For the latter one, additionally, the results of the deconvolution with the mapping matrix following \citet{plantier_etal_12} are given.
Interpolation smears the autocovariance function, yielding deviations, if estimated without the deconvolution (Fig.~\ref{fig:simmrand}a).
More obvious this dynamic error becomes with the autospectrum in Fig.~\ref{fig:simmrand}c, where a frequency dependent deviation from the reference spectrum occurs, which is characteristic for the interpolation scheme used.
Lomb-Scargle's method has a significant systematic error of the autocovariance estimate at lag time zero corresponding to an offset of the autospectrum.
Principally, this offset can be estimated and corrected using
$$\hat {\vecstyle{S}}=\vecstyle{S}-\frac{\Delta\tau}{D}\left(\frac{1}{\alpha'}-1\right)\sum\limits_{i=0}^{N-1} w_i \left(z_i-\bar z\right)^2$$
where $\vecstyle{S}$ is the primary, biased estimate of Lomb-Scargle's method, $\alpha'$ is the probability of a sample to be valid and $\hat {\vecstyle{S}}$ is the improved estimate of the power spectral density.
For details, see \citet{random_and_gappy_direct}.
%
%
However, this offset correction is suited only for randomly occurring and independent data points missing.
Interpolation with deconvolution and the estimators using only valid samples show no such significant deviations from the expected spectrum (Fig.~\ref{fig:simmrand}c).
However, for all estimators without the correction for short data sets, there remains a deviation of the autocovariance estimates (Fig.~\ref{fig:simmrand}a) if the estimated mean value is removed from the data.
The additional correction for short data sets is able to correct this error.
However, it is available only for the estimators using only valid samples.
With an equivalent correction, also sample-and-hold interpolation with deconvolution as well as Lomb-Scargle's method could obtain improved estimates of the autocovariance and the spectrum.
However, this holds only for strictly independent outliers, since the deconvolution procedure implies this assumption and also Lomb-Scargle's method has a significant dynamic error for correlated data gaps.

For the cross-covariance and the cross-spectral density the deconvolution after interpolation has not been derived by \citet{plantier_etal_12}.
Without this deconvolution, the dynamic error will always affect the results for interpolation leading to similar conclusions as for the autocovariance and spectrum.
No such deviation from the expected values can be identified for the new estimation methods except for the remaining deviation due to the short record length if no appropriate correction is used.
The additional correction for short data sets is able to correct this error also for the cross-covariance case.

The bias due to the short data sets mainly consists of an offset.
The linear term in the estimate of the cross-covariance becomes not obvious in the present simulation.
%
However, in other spectral compositions it becomes more obvious and may even lead to an over-estimation of the cross-covariance at certain lag times. 
The bias of the autocovariance function is symmetric and, therefore, has no linear term.
Higher-order contributions exist.
However, they are significantly smaller.

Due to the asymmetry of the cross-covariance function the cross-spectral density becomes complex.
In Fig.~\ref{fig:simmrand}d only the magnitude of the mean complex power spectral density is shown.
In the log-log plot of the spectra (both, auto and cross) only one half of the spectrum is shown and also the value at frequency zero is not shown.
Since the bias in the covariance functions mainly consists of an offset, this bias does not contribute significantly to the spectrum at frequencies different than zero.
Therefore, changes due to the bias-correction for short data sets contribute only little to the values of the spectra shown in the diagrams.

The correction will be complete only under the assumption, that no correlation exists outside the observed interval. 
Since in the present simulation, this is not entirely fulfilled, the correction is not exact.
However, the deviation from this assumption is small enough and therefore the correction is sufficient with no obvious bias remaining.
Furthermore, this evidences that the correction is robust against (small) remaining correlations outside the observed correlation interval.

\begin{figure}
\centering{%
\includegraphics[scale=0.79]{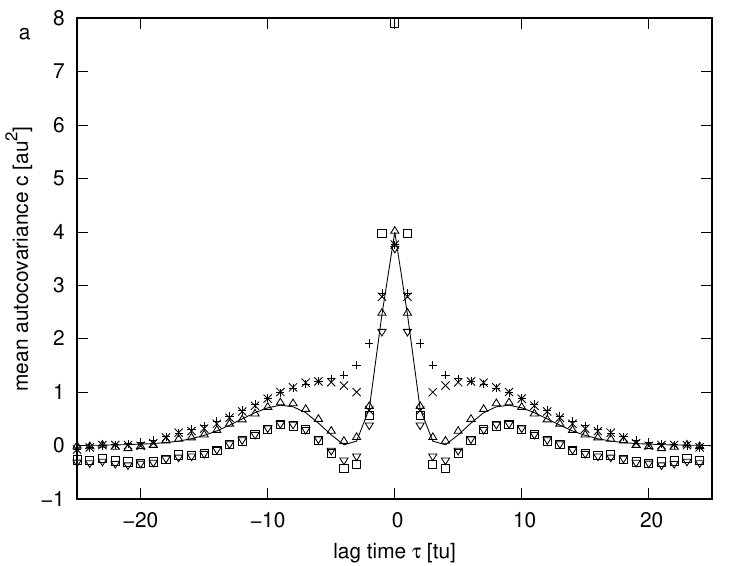}%
\hfill
\includegraphics[scale=0.79]{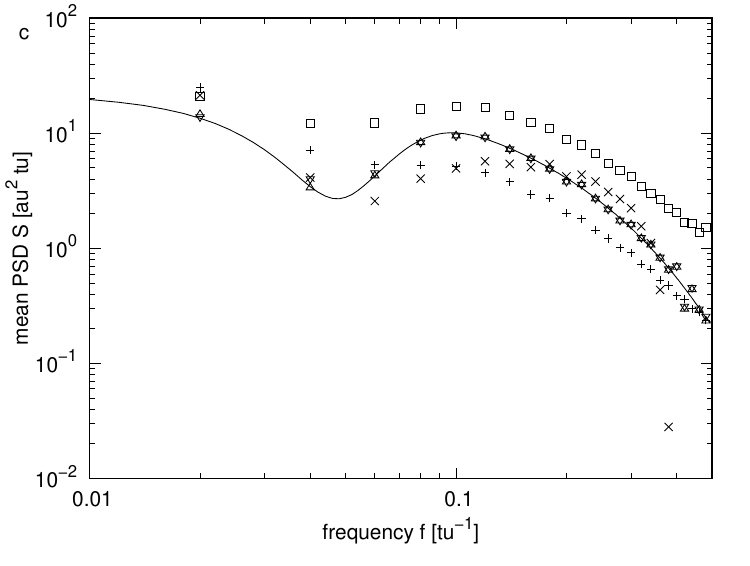}%
}\\
\centering{%
\includegraphics[scale=0.79]{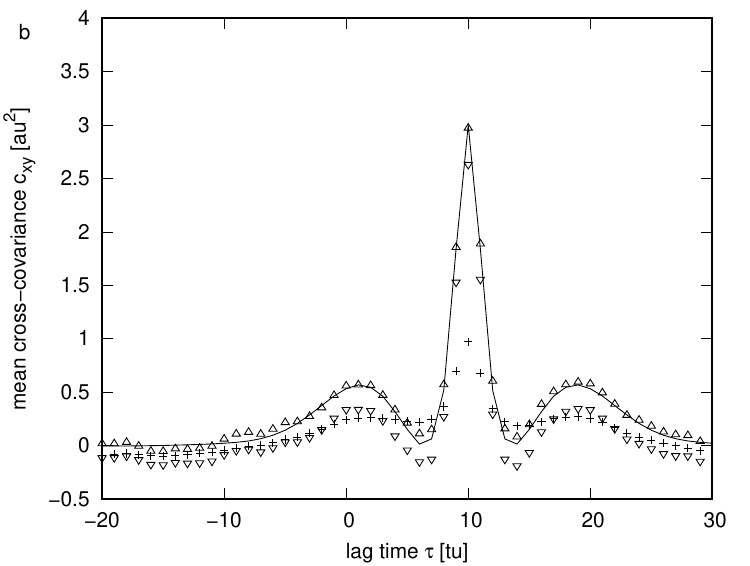}%
\hfill
\includegraphics[scale=0.79]{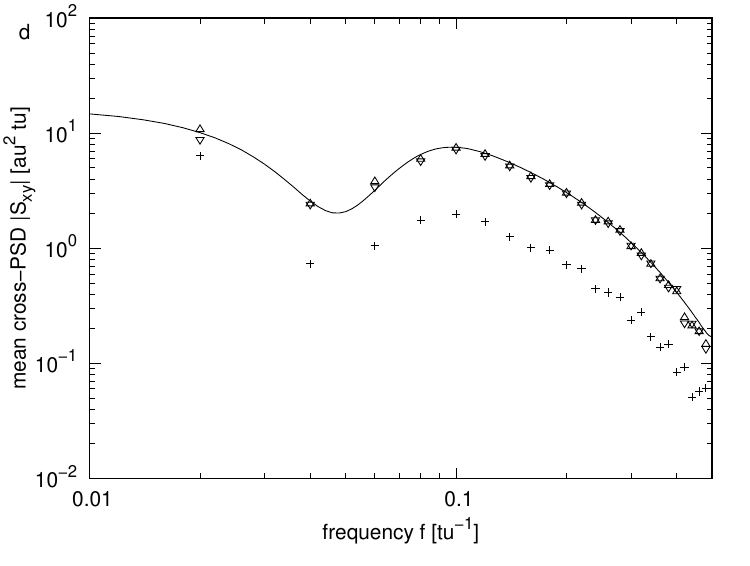}%
}
\caption{Empirical mean from averages over 1\,000 realizations with long data gaps and 100 samples each: 
(a) autocovariance function, (b) cross-covariance function, (c) power spectral density (autospectrum), (d) magnitude of the cross power spectral density (cross-spectrum,
$\unit{au}$ - amplitude unit, $\unit{tu}$ - time unit,
$+$ sample-and-hold interpolation w/o deconvolution, $\times$ sample-and-hold interpolation w deconvolution, $\square$ Lomb-Scargle's method, $\bigtriangledown$ only valid points w/o correction for short data sets, $\bigtriangleup$ only valid points w correction for short data sets, --- simulation).}
\label{fig:simmlong}
\end{figure}

Fig.~\ref{fig:simmlong} shows the mean values of the estimated functions for the simulation with long data gaps, where the occurrence of invalid samples is correlated.
In this case, the model for the deconvolution after sample-and-hold interpolation is not suited.
Therefore, both, the autocovariance function as well as the autospectral density (Fig.~\ref{fig:simmlong}a and c), show significant deviations for the interpolation method even with deconvolution.
The deconvolution is not able to compensate the dynamic error due to the violation of the assumptions for the derivation of the deconvolution matrix.
Without the deconvolution, interpolation always has such a dynamic error.
Lomb-Scargle's method also has such a dynamic error, as can be noticed in the autocorrelation function as an additional smearing of the peak around zero lag time.
Correspondingly, the error in the spectrum is not simply an offset any more and the above correction of the spectral estimate will fail here.
On the contrary, the methods based on the average over valid samples only, show no such dynamic errors and the correction for short data sets finally yields bias-free estimates.
This holds for both, the autocovariance and spectrum estimation as well as for the cross-covariance and spectrum.

\subsection{Root mean square error}
\label{sec:RMS}

With increasing record length, more and more pairs of valid samples of a certain lag time get available.
The averages converge accordingly and the estimator variance of the covariance function decreases for each and for all lag times. 
As long as the support of the covariance function considered for the transform into a spectrum remains unchanged, also the estimator variance of the spectrum decreases accordingly.
This also holds for bias correction for short data sets, since the correction consists of a linear transformation of the primary estimates.
Finally, one can conclude that the introduced method for bias-free estimation is consistent.
This holds independent of the spectral characteristics of the data gaps.
Therefore, it is not necessary to investigate random outliers and longer data gaps in comparison.
Since both cases are representative, from here on only the results for the more rich case of data sets with long data gaps are presented.
Furthermore, this is the case, where other methods potentially fail.

\begin{figure}
\centering{%
\includegraphics[scale=0.79]{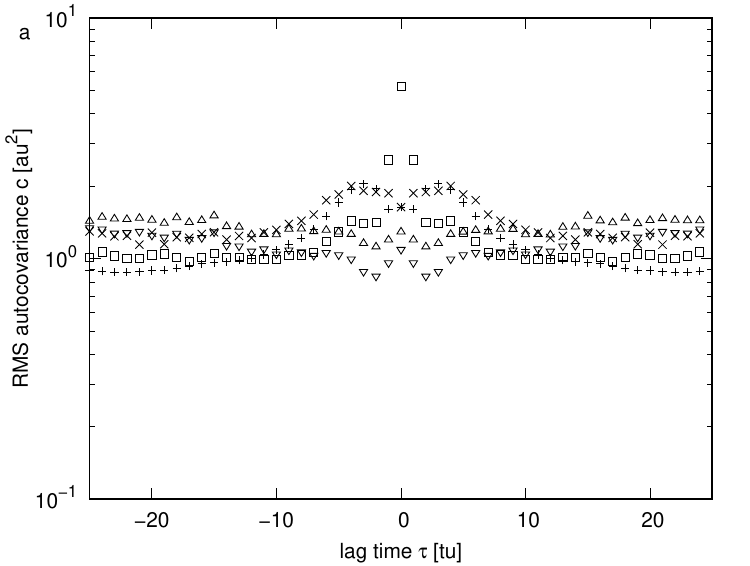}%
\hfill
\includegraphics[scale=0.79]{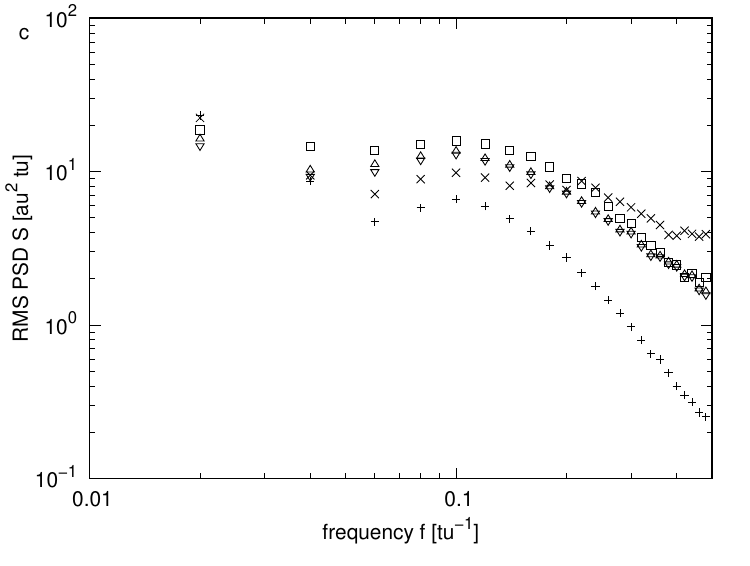}%
}\\
\centering{%
\includegraphics[scale=0.79]{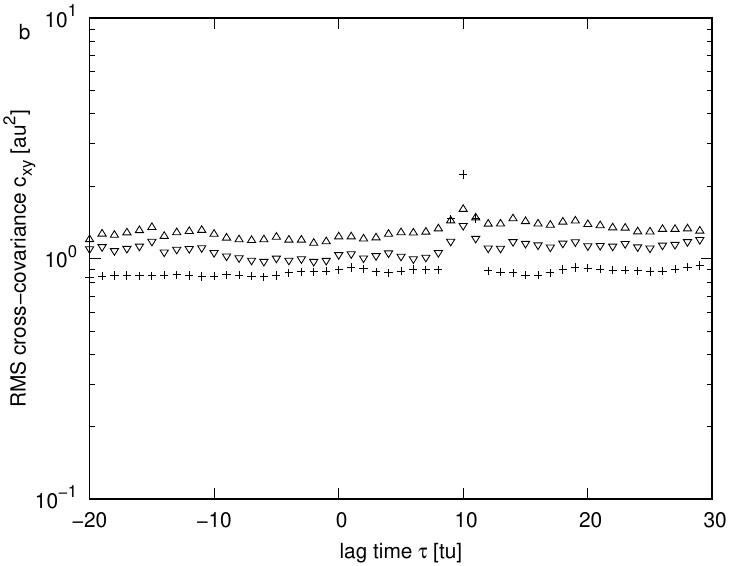}%
\hfill
\includegraphics[scale=0.79]{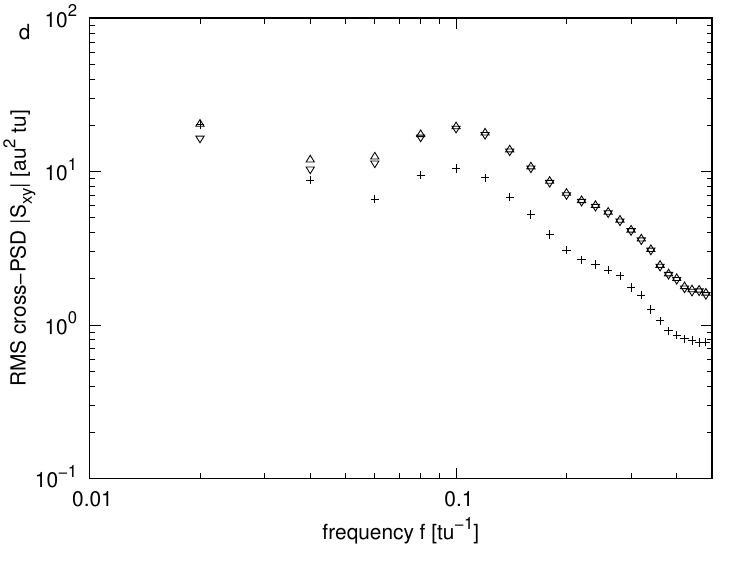}%
}
\caption{Empirical root mean square (RMS) error from averages over 1\,000 realizations with long data gaps and 100 samples each (valid and invalid samples together): 
(a) autocovariance function, (b) cross-covariance function, (c) power spectral density (autospectrum), (d) magnitude of the cross power spectral density (cross-spectrum,
$\unit{au}$ - amplitude unit, $\unit{tu}$ - time unit,
$+$ sample-and-hold interpolation w/o deconvolution, $\times$ sample-and-hold interpolation w deconvolution, $\square$ Lomb-Scargle's method, $\bigtriangledown$ only valid points w/o correction for short data sets, $\bigtriangleup$ only valid points w correction for short data sets).}
\label{fig:simrmsshort}
\end{figure}

Fig.~\ref{fig:simrmsshort} shows the empirical RMS errors obtained from the previous simulation with long data gaps.
Since the new estimator is bias-free, its RMS error can be smaller than that of the biased estimators where ever the latter ones have a strong enough bias.
In comparison to Lomb-Scarge's estimator this is obvious for the autocovariance estimate in Fig.~\ref{fig:simrmsshort}a for small lag times, where strong correlations exist and a strong bias has been found for Lomb-Scarge's estimator (cf.\ Figs.~\ref{fig:simmrand}a and \ref{fig:simmlong}a).
For all other lag times, the estimators all have about the same RMS errors with minor differences.
However, sample-and-hold interpolation without deconvolution owes its small RMS error only to the strong damping of this method, corresponding to its significant dynamic error (cf.\ Figs.~\ref{fig:simmrand}c and \ref{fig:simmlong}c).
Coincidentally, this yields values tending towards the correct ones since correlations are required to vanish at larger lag times for the correction procedure for short data sets.
The estimators with valid data points only have an almost constant RMS error for all lag times. 
Including correction for short data sets increases the RMS error slightly.
This also holds for the cross-correlation and cross-spectrum (Figs.~\ref{fig:simrmsshort}b and d). 
Due to damping, the biased estimates of the interpolation finally yield lower RMS errors than those of the bias-free estimators with valid samples only.

\begin{figure}
\centering{%
\includegraphics[scale=0.79]{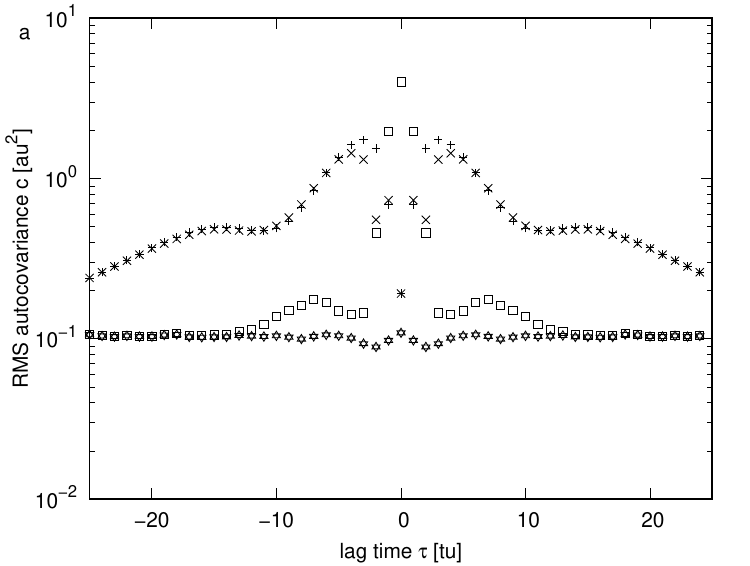}%
\hfill
\includegraphics[scale=0.79]{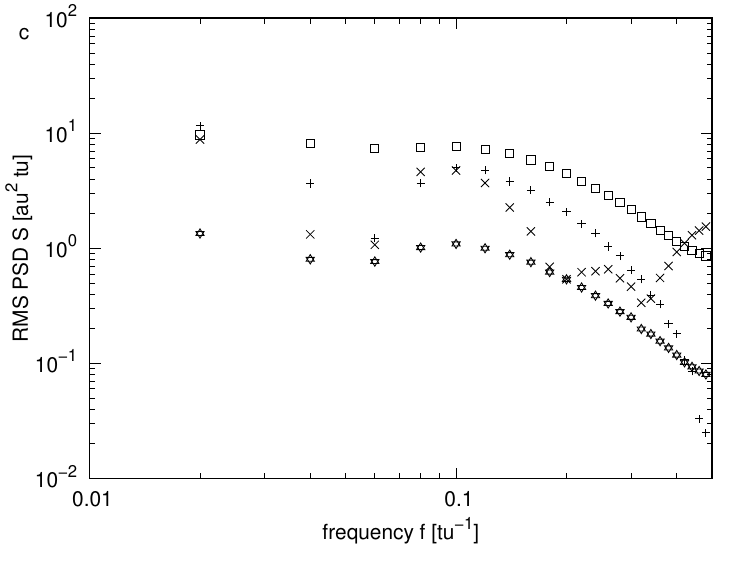}%
}\\
\centering{%
\includegraphics[scale=0.79]{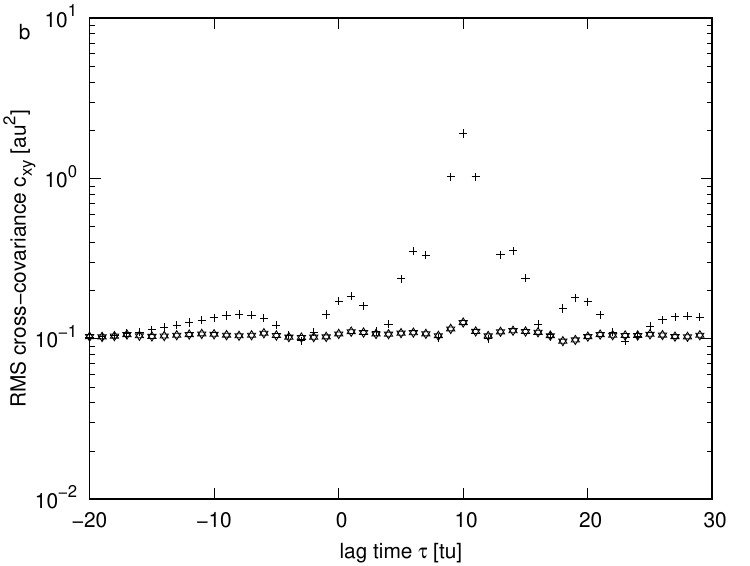}%
\hfill
\includegraphics[scale=0.79]{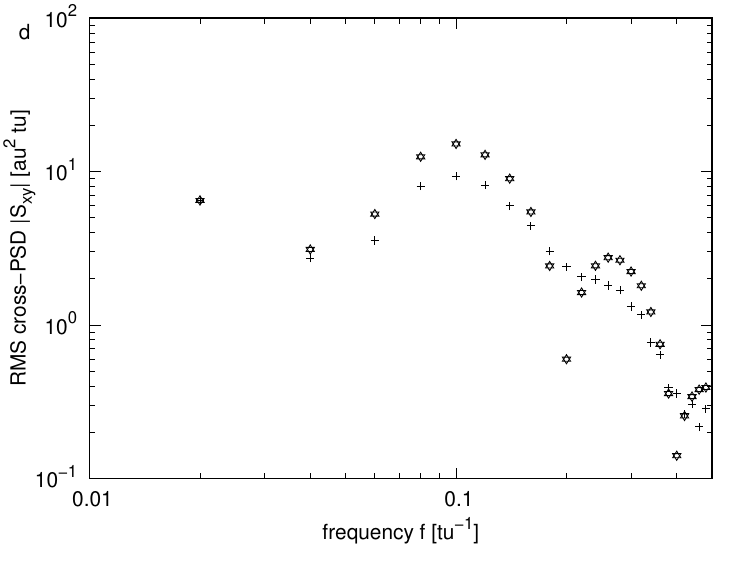}%
}
\caption{Empirical root mean square (RMS) error from averages over 1\,000 realizations with long data gaps and 10\,000 samples each (valid and invalid samples together): 
(a) autocovariance function, (b) cross-covariance function, (c) power spectral density (autospectrum), (d) magnitude of the cross power spectral density (cross-spectrum,
$\unit{au}$ - amplitude unit, $\unit{tu}$ - time unit,
$+$ sample-and-hold interpolation w/o deconvolution, $\times$ sample-and-hold interpolation w deconvolution, $\square$ Lomb-Scargle's method, $\bigtriangledown$ only valid points w/o correction for short data sets, $\bigtriangleup$ only valid points w correction for short data sets).}
\label{fig:simrmslonglong}
\end{figure}

The advantage of bias-free estimation becomes obvious only for longer data records, especially for correlated data gaps, as shown in Fig.~\ref{fig:simrmslonglong} for data sets with 10\,000 samples each.
All other simulation parameters have held unchanged.
While bias-free estimators with increasing amount of data converge towards the correct values yielding smaller and finally vanishing RMS errors, biased estimators cannot fall below a certain value.
Therefore, the RMS errors of the new estimators easily fall below the RMS errors of the other (biased) estimators.
This holds at least at those lag times, where these estimators yield biased estimates, while Lomb-Scargle's method still has low RMS errors at large lag times.
However, the RMS error of the spectrum (Fig.~\ref{fig:simrmslonglong}c) uniquely demonstrates the advantage of the new methods using the valid samples only. 
The absence of any bias for these methods finally also leads to smaller RMS errors compared to biased estimators, at least for an increasing amount of data.

\section{Conclusion}

Routines for non-parametric estimation of the mean value, the variance, the auto- and cross-covariance function as well as the corresponding power spectral densities from signals with invalid samples have been introduced.
The information about the validity of the samples is assumed {\em a priori} known.
The introduced estimation methods are based on ensemble averages over valid samples only, allowing bias-free estimates.
Both sources of systematic errors are tackled, data gaps as well as too short data sets.
The routines allow bias-free estimation, independent of the dynamic characteristics of the gaps. 
%
%
%
Furthermore, the estimators are consistent, yielding vanishing RMS errors for increasing amounts of data.
This is the advantage compared to other, biased estimates, which on the contrary have remaining RMS errors even for infinite an amount of data.

For independent outliers, Lomb-Scargle's method yields a significant offset of the spectral estimate. 
For correlated data gaps, a dynamic error adds.
Interpolation without deconvolution shows such a dynamic error independent of the spectral composition of the sampling scheme.
Dynamic errors can miss-lead the interpretation of the process under investigation.
Appropriate deconvolution can invert and correct this dynamic error.
However, detailed information is required about the sampling scheme in correspondence with the interpolation.
So far, appropriate deconvolution matrices are available only for simple case like purely random outliers with sample-and-hold interpolation.
For any other combination (different interpolation or correlated sampling schemes), the available deconvolution procedures are not suited and need significant adaptation.

Bias-free estimation, especially {\em a posteriori} bias corrections, typically come with an increase of estimation uncertainties.
Increasing the amount of data can efficiently reduce estimation uncertainties.
Appropriate RMS errors vanish due to the absence of systematic errors.
Therefore, bias-free estimation outplays its advantages tending for longer data sets.
For those, the bias-correction from Sec.~\ref{sec:short} can be omitted, because the simple averages from Sec.~\ref{sec:gappy} are asymptotically bias-free and for long enough data sets they are sufficient.
However, bias correction for short data sets can be applied by default also to longer data sets to ensure bias-free estimates independent from the amount of data.
There is no risk of an overcorrection or making the bias worse.
%
%
%
%
%
%
%
In the present study, the procedures for bias-free estimation have been applied to gappy data sets with weights of one for valid samples and zero for invalid samples.
However, the procedures including the correction for short data sets are suitable also for other weights, including non-binary values.
The programs used here are available with the present article as supplementary material.

\appendix

\section{Derivation of Eqs.~(\ref{eq:esq}) and (\ref{eq:varicorrect})}
\label{sec:appexpvari}

\noindent
From Eq.~(\ref{eq:vari}) follows
\begin{equation}
s^2=\frac{1}{D}\left(\sum\limits_{i=0}^{N-1} w_i z_i^2\right)
-\frac{2}{D}\left(\sum\limits_{i=0}^{N-1} w_i z_i\bar z\right)+\frac{1}{D}\left(\sum\limits_{i=0}^{N-1} w_i \bar z^2\right)
\end{equation}
\begin{eqnarray}
&=&\frac{1}{D}\left(\sum\limits_{i=0}^{N-1} w_i z_i^2\right)-\frac{2}{D}\left\{\sum\limits_{i=0}^{N-1} w_i z_i \left[\frac{1}{D}\left(\sum\limits_{j=0}^{N-1} w_j z_j\right)\right]\right\}\nonumber\\
&&+\frac{1}{D}\left\{\sum\limits_{i=0}^{N-1} w_i \left[\frac{1}{D}\left(\sum\limits_{j=0}^{N-1} w_j z_j\right)\right]^2\right\}
\end{eqnarray}
\begin{eqnarray}
&=&\frac{1}{D}\left(\sum\limits_{i=0}^{N-1} w_i z_i^2\right)
-\frac{2}{D^2}\left(\sum\limits_{i=0}^{N-1}\sum\limits_{j=0}^{N-1} w_i w_j z_i z_j\right)+\left[\frac{1}{D}\left(\sum\limits_{j=0}^{N-1} w_j z_j\right)\right]^2
\end{eqnarray}
\begin{eqnarray}
&=&\frac{1}{D}\left(\sum\limits_{i=0}^{N-1} w_i z_i^2\right)
-\frac{2}{D^2}\left(\sum\limits_{i=0}^{N-1}\sum\limits_{j=0}^{N-1} w_i w_j z_i z_j\right)+\frac{1}{D^2}\left(\sum\limits_{i=0}^{N-1}\sum\limits_{j=0}^{N-1} w_i w_j z_i z_j\right)\nonumber\\
\end{eqnarray}
\begin{eqnarray}
&=&\frac{1}{D}\left(\sum\limits_{i=0}^{N-1} w_i z_i^2\right)
-\frac{1}{D^2}\left(\sum\limits_{i=0}^{N-1}\sum\limits_{j=0}^{N-1} w_i w_j z_i z_j\right)
\end{eqnarray}
\begin{eqnarray}
&=&\frac{1}{D^2}\left[\sum\limits_{i=0}^{N-1}\sum\limits_{j=0}^{N-1} w_i w_j \left( z_i^2 - z_i z_j\right)\right].
\end{eqnarray}

\noindent
The expectation of $s^2$ then is
\begin{equation}
\langle s^2\rangle =\frac{1}{D^2}\left\{\sum\limits_{i=0}^{N-1}\sum\limits_{j=0}^{N-1} w_i w_j \left[ \left( \sigma^2 +\mu^2\right)- \left( \gamma_{j-i}+\mu^2\right)\right]\right\}
\end{equation}
\begin{equation}
=\frac{1}{D^2}\left(\sum\limits_{i=0}^{N-1}\sum\limits_{j=0}^{N-1} w_i w_j \sigma^2\right)-\frac{1}{D^2}\left(\sum\limits_{i=0}^{N-1}\sum\limits_{j=0}^{N-1} w_i w_j \gamma_{j-i}\right)
\end{equation}
\begin{equation}
=\sigma^2-\sigma_{\bar z}^2
\end{equation}

\noindent
Finally, $s^2+\sigma_{\bar z}^2$ is a bias-free estimate of the variance $\sigma^2$.

\section{Derivation of Eqs.~(\ref{eq:eck}) and (\ref{eq:eek})}
\label{sec:appcov}

\noindent
From Eq.~(\ref{eq:autocovariance}) follows
\begin{eqnarray}
C_k&=&\frac{1}{W_k}\left[\sum\limits_{i=I_1}^{I_2} w_i w_{i+k} z_i z_{i+k}\right]-\frac{\bar z}{W_k}\left[\sum\limits_{i=I_1}^{I_2} w_i w_{i+k} \left(z_i+z_{i+k}\right)\right]+\bar z^2\qquad
\end{eqnarray}
\begin{eqnarray}
&=&\frac{1}{W_k}\left[\sum\limits_{i=I_1}^{I_2} w_i w_{i+k} z_i z_{i+k}\right]-\frac{1}{D W_k}\left(\sum\limits_{j=0}^{N-1} w_j z_j\right)\left[\sum\limits_{i=N_1}^{N_2} w_i w_{i+k} \left(z_i+z_{i+k}\right)\right]\nonumber\\
&&+\left[\frac{1}{D}\left(\sum\limits_{j=0}^{N-1} w_j z_j\right)\right]^2
\end{eqnarray}
\begin{eqnarray}
&=&\frac{1}{W_k}\left[\sum\limits_{i=I_1}^{I_2} w_i w_{i+k} z_i z_{i+k}\right]-\frac{1}{D W_k}\left[\sum\limits_{i=I_1}^{I_2} \sum\limits_{j=0}^{N-1} w_i w_{i+k} w_j \left(z_i+z_{i+k}\right)z_j\right]\nonumber\\
&&+\frac{1}{D^2} \left(\sum\limits_{i=0}^{N-1} \sum\limits_{j=0}^{N-1} w_i w_j z_i z_j\right)
\end{eqnarray}
with $I_1=\max(0,-k)$ and $I_2=\min(N,N-k)-1$.

\noindent
The expectation of $C_k$ then is
\begin{eqnarray}
\langle C_k\rangle &=&\frac{1}{W_k}\left[\sum\limits_{i=I_1}^{I_2} w_i w_{i+k} \left(\gamma_k+\mu^2\right)\right]
\nonumber\\
&&-\frac{1}{D W_k}\left[\sum\limits_{i=I_1}^{I_2} \sum\limits_{j=0}^{N-1} w_i w_{i+k} w_j \left(\gamma_{j-i}+\gamma_{j-(i+k)}+2\mu^2\right)\right]
\nonumber\\
&&+\frac{1}{D^2}\left[\sum\limits_{i=0}^{N-1} \sum\limits_{j=0}^{N-1} w_i w_j \left(\gamma_{j-i}+\mu^2\right)\right]
\end{eqnarray}
\begin{eqnarray}
&=&\gamma_k
+\frac{1}{D^2}\left(\sum\limits_{i=0}^{N-1} \sum\limits_{j=0}^{N-1} w_i w_j \gamma_{j-i}\right)
\nonumber\\
&&-\frac{1}{D W_k}\left[\sum\limits_{i=I_1}^{I_2} \sum\limits_{j=0}^{N-1} w_i w_{i+k} w_j \left(\gamma_{j-i}+\gamma_{i+k-j}\right)\right]
\end{eqnarray}
\begin{eqnarray}
&=&\gamma_k
+\sigma_{\bar z}^2-\frac{1}{D W_k}\left[\sum\limits_{i=I_1}^{I_2} \sum\limits_{j=0}^{N-1} w_i w_{i+k} w_j \left(\gamma_{j-i}+\gamma_{i+k-j}\right)\right].\quad
\end{eqnarray}

\section{Derivation of Eqs.~(\ref{eq:eckx}) and (\ref{eq:eekx})}
\label{sec:appcovx}

\noindent
From Eq.~(\ref{eq:crosscovariance}) follows
\begin{eqnarray}
C_{\rmxy,k}&=&\frac{1}{W_{\rmxy,k}}\left[\sum\limits_{i=N_1}^{N_2} w_{\rmx,i} w_{\rmy,i+k} z_{\rmx,i} z_{\rmy,i+k}\right]-\frac{\bar z_\rmy}{W_{\rmxy,k}}\left[\sum\limits_{i=N_1}^{N_2} w_{\rmx,i} w_{\rmy,i+k} z_{\rmx,i}\right]
\nonumber\\
&&-\frac{\bar z_\rmx}{W_{\rmxy,k}}\left[\sum\limits_{i=N_1}^{N_2} w_{\rmx,i} w_{\rmy,i+k} z_{\rmy,i+k}\right]
+\bar z_\rmx \bar z_\rmy
\end{eqnarray}
\begin{eqnarray}
&=&\frac{1}{W_{\rmxy,k}}\left[\sum\limits_{i=N_1}^{N_2} w_{\rmx,i} w_{\rmy,i+k} z_{\rmx,i} z_{\rmy,i+k}\right]
\nonumber\\
&&-\frac{1}{D_\rmy W_{\rmxy,k}}\left(\sum\limits_{j=0}^{N_\rmy-1} w_{\rmy,j} z_{\rmy,j}\right)
\left[\sum\limits_{i=N_1}^{N_2} w_{\rmx,i} w_{\rmy,i+k} z_{\rmx,i}\right]\nonumber\\
&&-\frac{1}{D_\rmx W_{\rmxy,k}}\left(\sum\limits_{j=0}^{N_\rmx-1} w_{\rmx,j} z_{\rmx,j}\right)
\left[\sum\limits_{i=N_1}^{N_2} w_{\rmx,i} w_{\rmy,i+k} z_{\rmy,i+k}\right]\nonumber\\
&&+\frac{1}{D_\rmx D_\rmy}\left(\sum\limits_{j=0}^{N_\rmx-1} w_{\rmx,j} z_{\rmx,j}\right)\left(\sum\limits_{j=0}^{N_\rmy-1} w_{\rmy,j} z_{\rmy,j}\right)
\end{eqnarray}
\begin{eqnarray}
&=&\frac{1}{W_{\rmxy,k}}\left[\sum\limits_{i=N_1}^{N_2} w_{\rmx,i} w_{\rmy,i+k} z_{\rmx,i} z_{\rmy,i+k}\right]
\nonumber\\
&&-\frac{1}{D_\rmy W_{\rmxy,k}}\left[\sum\limits_{i=N_1}^{N_2} \sum\limits_{j=0}^{N_\rmy-1} w_{\rmx,i} w_{\rmy,i+k} w_{\rmy,j} z_{\rmx,i} z_{\rmy,j}\right]
\nonumber\\
&&-\frac{1}{D_\rmx W_{\rmxy,k}}\left[\sum\limits_{i=N_1}^{N_2} \sum\limits_{j=0}^{N_\rmx-1} w_{\rmx,i} w_{\rmy,i+k} w_{\rmx,j} z_{\rmy,i+k} z_{\rmx,j}\right]
\nonumber\\
&&+\frac{1}{D_\rmx D_\rmy}\left(\sum\limits_{i=0}^{N_\rmx-1} \sum\limits_{j=0}^{N_\rmy-1} w_{\rmx,i} w_{\rmy,j} z_{\rmx,i} z_{\rmy,j}\right)
\end{eqnarray}
with $I_1=\max(0,-k)$ and $I_2=\min(N_\rmx,N_\rmy-k)-1$.

\noindent
The expectation of $C_{\rmxy,k}$ then is
\begin{eqnarray}
\langle C_{\rmxy,k}\rangle &=&\frac{1}{W_{\rmxy,k}}\left[\sum\limits_{i=N_1}^{N_2} w_{\rmx,i} w_{\rmy,i+k} \left(\gamma_{\rmxy,k}+\mu_\rmx \mu_\rmy\right)\right]
\nonumber\\
&&-\frac{1}{D_\rmy W_{\rmxy,k}}\left[\sum\limits_{i=N_1}^{N_2} \sum\limits_{j=0}^{N_\rmy-1} w_{\rmx,i} w_{\rmy,i+k} w_{\rmy,j} \left(\gamma_{\rmxy,j-i}+\mu_\rmx \mu_\rmy\right)\right]
\nonumber\\
&&-\frac{1}{D_\rmx W_{\rmxy,k}}\left[\sum\limits_{i=N_1}^{N_2} \sum\limits_{j=0}^{N_\rmx-1} w_{\rmx,i} w_{\rmy,i+k} w_{\rmx,j} \left(\gamma_{\rmxy,i+k-j}+\mu_\rmx \mu_\rmy\right)\right]
\nonumber\\
&&+\frac{1}{D_\rmx D_\rmy}\left[\sum\limits_{i=0}^{N_\rmx-1} \sum\limits_{j=0}^{N_\rmy-1} w_{\rmx,i} w_{\rmy,j} \left(\gamma_{\rmxy,j-i}+\mu_\rmx \mu_\rmy\right)\right]
\end{eqnarray}
\begin{eqnarray}
&=&\gamma_{\rmxy,k}
+\frac{1}{D_\rmx D_\rmy}\left(\sum\limits_{i=0}^{N_\rmx-1} \sum\limits_{j=0}^{N_\rmy-1} w_{\rmx,i} w_{\rmy,j} \gamma_{\rmxy,j-i}\right)
\nonumber\\
&&-\frac{1}{D_\rmy W_{\rmxy,k}}\left[\sum\limits_{i=N_1}^{N_2} \sum\limits_{j=0}^{N_\rmy-1} w_{\rmx,i} w_{\rmy,i+k} w_{\rmy,j} \gamma_{\rmxy,j-i}\right]
\nonumber\\
&&-\frac{1}{D_\rmx W_{\rmxy,k}}\left[\sum\limits_{i=N_1}^{N_2} \sum\limits_{j=0}^{N_\rmx-1} w_{\rmx,i} w_{\rmy,i+k} w_{\rmx,j} \gamma_{\rmxy,i+k-j}\right].\qquad
\end{eqnarray}

\bibliography{Gappy_BiasFree_arXiv}

\end{document}